\documentclass[twocolumn]{aastex62}

\received{\bf October 17, 2018}
\revised{\bf January 24, 2019}
\accepted{\bf  January 24, 2019}
\submitjournal{ApJ}

\shorttitle{Evaluating the  classification of $Fermi$ BCUs }  
\shortauthors{Kang et al.}

\begin{document}

\title{Evaluating the optical classification of Fermi BCUs using machine learning}

\correspondingauthor{Shi-Ju Kang}\email{kangshiju@hust.edu.cn, kangshiju@alumni.hust.edu.cn}

\author[0000-0002-9071-5469]{Shi-Ju Kang}
\affil{School of Electrical Engineering, Liupanshui Normal University, Liupanshui, Guizhou, 553004, China}
\affiliation{Guizhou Provincial Key Laboratory of Radio Astronomy and Data Processing}

\author{Jun-Hui Fan}
\affiliation{Center for Astrophysics, Guangzhou University, Guangzhou 510006, China}

\author{Weiming Mao}
\affiliation{Department of Physics, Yunnan Normal University, Kunming, Yunnan, 650092, China}

\author[0000-0003-4773-4987]{Qingwen Wu}
\affiliation{School of Physics, Huazhong University of Science and Technology, Wuhan, Hubei, 430074, China}
\collaboration{}

\author{Jianchao Feng}
\affiliation{School of Physics and Electronic Science, Guizhou Normal University, Guiyang,  550001, China}
\affiliation{Guizhou Provincial Key Laboratory of Radio Astronomy and Data Processing}

\author{Yue Yin}
\affiliation{School of Electrical Engineering, Liupanshui Normal University, Liupanshui, Guizhou, 553004, China}



\begin{abstract}

In  the third catalog of active galactic nuclei detected by the $Fermi$-LAT (3LAC) Clean Sample, there are 402 blazars candidates of uncertain type (BCU). {Due to the limitations of astronomical observation or intrinsic properties, it is difficult to classify blazars using optical spectroscopy.} {The potential classification of BCUs using machine learning algorithms is essential.} Based on the 3LAC Clean Sample,  we collect 1420 Fermi blazars with 8 parameters of $\gamma$-ray photon spectral index, radio flux, flux density, curve significance, the integral  photon flux in 100 to 300 MeV,  0.3 to 1 GeV, 10 to 100 GeV  and variability index. {Here, we apply 4 different supervised machine learning (SML) algorithms} (\emph{Decision trees, Random forests, support vector machines and $Mclust$ Gaussian finite mixture models}) to evaluate the classification of {BCUs} based on the direct observational properties. All the 4 methods can perform exceedingly well with a more accuracy and can effective forecast the classification of $Fermi$ {BCUs}. The evaluating results show the results of these methods {(SML)} are valid and robust,  where, about 1/4 sources are FSRQs and 3/4 are BL Lacs  in 400 {BCUs}, which are consistent with some other recent results. {Although a} number of factors influence the accuracy of {SML,} the results are stable at a fixed ratio 1:3 between FSRQs and BL Lacs{, which} suggests that the {SML} can provides an effective method to evaluate the  {potential} classification of {BCUs}. {Among the 4 methods,} $Mclust$ Gaussian Mixture Modelling {has the highest} accuracy for our training sample (4/5, seed=123).

\end{abstract}



\section{Introduction} \label{sec:intro}

Blazars are a peculiar sub-class of radio-loud active galactic nuclei (AGNs), whose broadband emission is mainly dominated by non-thermal components produced in a relativistic jet pointed at a small viewing angle to the line of sight \citep{1995PASP..107..803U}. According to the features of optical emission-line in blazars,  they are traditionally sub-divided into two groups: flat spectrum radio quasars (FSRQs) and BL Lacertae objects (BL Lacs), where the BL Lacs have weak or no emission lines (e.g., equivalent width, EW, of the emission line in rest frame is less than ${\rm 5 \AA}$) while FSRQs show stronger emission lines (${\rm EW\geq5 \AA}$) \citep{1995PASP..107..803U,1991ApJS...76..813S,1991ApJ...374..431S} in their optical spectra. The multi-wavelength spectral energy distributions (SEDs) from radio to  $\gamma$-ray bands of blazars dominantly comes from the non-thermal emission, where the SED normally exhibits a two-hump structure in the ${\rm log \nu-log \nu F_{\nu}}$ space. The lower energy hump (peaked at between millimeter and soft X-ray waveband) is normally attributed to the synchrotron emission produced by the non-thermal electrons in the jet, while the second hump (peaked at the MeV-GeV range) mainly comes from inverse Compton (IC) scattering.  {The location} of the peak for the lower energy bump in the SED, $\nu^{\rm S}_{\rm p}$, is used to classify the sources as low (LSP, e.g., $\nu^{\rm S}_{\rm p}<10^{14}$ Hz), intermediate (ISP, e.g., $10^{14}~\rm Hz<\nu^{\rm S}_{\rm p}<10^{15}$ Hz) and high-synchrotron-peaked (HSP, e.g., $\nu^{\rm S}_{\rm p}>10^{15}$ Hz) blazars \citep{2010ApJ...716...30A}.

{In 2015, the $Fermi$-LAT Third Source Catalog (3FGL) was publicly released \citep{2015ApJS..218...23A}.} The 3FGL catalog includes 3033 $\gamma$-ray sources: 2192 high-latitude {($|b|>10\arcdeg$)} and 841 low-latitude {($|b| \leq10\arcdeg$)} $\gamma$-ray sources, where most sources belong to blazars \citep{2015ApJ...810...14A}. Based on the 3FGL \citep{2015ApJS..218...23A}, the third catalog of AGNs detected by the $Fermi$-LAT (3LAC) {was} presented by \cite{2015ApJ...810...14A}. The high-confidence clean sample of the 3LAC (3LAC Clean Sample), using the first four years of the $Fermi$-LAT data, lists 1444 $\gamma$-ray AGNs \citep{2015ApJ...810...14A}, which include 414 FSRQs ($\sim$~30\%), 604 BL Lac objects ($\sim$~40\%), 402 blazar candidates of uncertain type (BCU, $\sim$~30\%) and 24 non-blazar AGNs ($<$~2\%).

{Classified FSRQs and BL Lacs are sources} with their optical classifications can be well identified from the literature and/or optical spectrum in the 3FGL catalog \citep{2015ApJ...810...14A,2015ApJS..218...23A}. BCUs are  the sources with their counterparts have been established. However, their optical classifications have not been identified as a FSRQ or a BL Lac from the weaker or lacking an optical spectrum, and/or their synchrotron peak frequencies of SED, and/or their broadband emission shows blazar-type characteristics with a flat radio spectrum (see \citealt{2015ApJ...810...14A,2015ApJS..218...23A} for the details and references therein). Such a large sample of blazars provides a good chance to explore the nature of $\gamma$-ray emission of blazars 
{(e.g., 
\citealt{2012ApJ...753...45S,
2014MNRAS.441.3375X,
2015MNRAS.454..115S,
2015MNRAS.451.2750X,
2015MNRAS.450.3568X,
2016RAA....16...13C,
2016ApJS..226...20F,
2016RAA....16..173F,
2016Galax...4...36G,
2016RAA....16..103L,
2017RAA....17...66L, 
2018ApJS..235...39C,
2018RAA....18...56K,
2018RAA....18..120L}).}
{In the 3LAC} Clean Sample, there are about 30\% of blazars (BCUs) that have no optical classification. Evaluating {potential} classification of the BCUs is  a meaningful topics, which have been extensively explored based on the Fermi source catalogs (e.g., see 
{
\citealt{2013MNRAS.428..220H,
2014ApJ...782...41D,
2016MNRAS.462.3180C,
2016Galax...4...14E,
2016ApJ...820....8S,
2017A&A...602A..86L,
2017MNRAS.470.1291S,
2017ApJ...838...34Y} for the reviews and references therein). 
}

At present, machine learning and data mining techniques are developing rapidly, 
which has been widely used in the study of astronomy and astrophysics 
(e.g., see the review in \citealt{2010IJMPD..19.1049B}; \citealt{feigelson_babu_2012}
and \citealt{2012amld.book.....W};
also see 
\citealt{2012ApJ...753...83A,2012MNRAS.424L..64M,
2013MNRAS.428..220H,
2014ApJ...782...41D,
2016MNRAS.462.3180C,
2016Galax...4...14E,
2016ApJ...820....8S,
2017A&A...602A..86L,
2017MNRAS.470.1291S,
2017ApJ...838...34Y} ;
\citealt{2018RAA....18..118B};
\citealt{2018arXiv181207190M}).
{Supervised machine learning (SML)} 
is  the most common technique, 
which aim is to build a classifier (or a decision rule) from the observations of known classification, 
to classify others (an observation with an unknown class membership to one of K known classes).
{In the 3LAC} Clean Sample, there are about 70\% sources 
{with known} optical classification (414 FSRQs $\sim$~30\%, 604 BL Lacs $\sim$~40\%),
however, there are about 30\% of blazars (BCUs) that have no optical classification.
Evaluating 
{potential} classification of the BCUs using 
{supervised machine learning} is an interesting work.

In this work,
{we employ 4 supervised machine learning algorithms} (\emph{Decision trees (DT), Random forests (RF), support vector machines (SVMs) and $Mclust$ Gaussian finite mixture models (Mclust)}) to evaluate the 
{potential} classification of BCUs only based on (only focus on) the direct observational properties of the 3LAC Clean Sample. 
We give some description on the sample selection in Section 2, and the
{supervised machine learning} techniques are introducted in Section 3. 
Section 4 reports the results of
{supervised machine learning.}
The discussion  and conclusion are presented in Section 5.

\begin{table*}[ht]
	\centering
	\caption{The result of two sample test for 604 BL Lacs  and 414 FSRQs}
	\label{tab:test}
	\begin{tabular}{clcccccccccccc} 
		\hline\hline
	&Selected                                     &\multicolumn{2}{c}{{KS test}}   & &\multicolumn{3}{c}{t-test} & &\multicolumn{2}{c}{Wilcox-test}    \\
\cline{3-4} \cline{6-8}	 \cline{10-11}			
{}&{Paramaters} & {$D$} & {$p_{1}$}&~& {$t$} &df& {$p_{2}$} &~& {$W$} & {$p_{3}$} &&{$Gini$} \\
		\hline
&	Spectral.Index	&	0.627	&	0.0	&	&	-27.096	&	955.17	&	2.10E-120	&	&	27783.0	&	7.42E-99	&&	90.89	\\
&	flux.density	&	0.627	&	0.0	&	&	-5.034	&	425.25	&	7.11E-07	&	&	33001.5	&	9.89E-89	&&	74.98 	\\
&	Radio.flux.mJy.	&	0.562	&	0.0	&	&	-5.659	&	425.62	&	2.80E-08	&	&	39308.0	&	3.08E-77	&&	55.55 	\\
&	variability.index	&	0.478	&	0.0	&	&	-3.096	&	419.60	&	2.10E-03	&	&	52423.0	&	6.24E-56	&&	45.65	\\
&	flux.100.300.mev&    0.472	&	0.0	&	&	-5.318	&	430.82	&	1.69E-07	&	&	50263.0	&	3.37E-59	&&	32.89 	\\
&	flux.0p3.1.gev	&	0.424	&	0.0	&	&	-4.559	&	443.64	&	6.67E-06	&	&	59379.5	&	4.72E-46	&&	24.04 	\\
&	flux.10.100.gev	&	0.405	&	0.0	&	&	5.874	&	818.82	&	6.17E-09	&	&	186303.0	&	2.40E-40	&&	47.94	\\
&	curve.significance&	0.274	&	2.22E-16	&	&-9.604	&	555.30	&	2.63E-20	&	&	84223.5	&	8.37E-19	&&	22.20	\\ 	
	\hline
	\end{tabular}
	\\
\tablecomments{Column 1 shows the parameters selected in sample; 
Column 2 and Column 3 give the value of the test statistic ($D$) and  p-value ($p_1$) for the two-sample Kolmogorov$-$Smirnov test;
The value of the t-statistic ($t$), the degrees of freedom for the t-statistic (df) and the p-value ($p_2$) for the Welch Two Sample t-test are listed in Column 4, Column 5  and Column 6 respectively; 
Column 7 and Column 8 report the value of the test statistic ($W$) and  p-value ($p_3$) for the Wilcoxon rank sum test with continuity correction; 
Column 9 lists the Gini coefficient ($Gini$) that is a natural measure of variable importance in Random forests. 
}
\end{table*}

\section{Sample} \label{sec:sample}

From the 3FGL catalog \citep{2015ApJS..218...23A} and 3LAC Clean catalog \citep{2015ApJ...810...14A}, 
we select 1420 Fermi Clean blazars (including 414 FSRQs, 604 BL Lacs  and 402 BCUs) 
{with 37 variables.}
{In order to select} suitable parameters for supervised machine-learning, and to built an available supervised classifier,
the independence of these 37 parameters distributions between two subsamples (414 FSRQs and 604 BL Lacs) are calculated using two sample test (KS test, t-test and Wilcox-test) 
{(e.g., \citealt{2018MNRAS.475.1708A})}.
Based on the two sample test results (see Table \ref{tab:test}), excluding the same, similar, and related parameters, 
or some parameters that are directly related to classification (e.g., redshift), 
{8 parameters} (the $\gamma$-ray photon spectral index ($\Gamma_{\rm ph}$), radio flux  (log${\rm F_R}$), flux density (log$F_D$), curve significance ($C_S$),
the integral  photon flux in 100 to 300 MeV (log$F_1$),  0.3 to 1 GeV (log$F_2$), 10 to 100 GeV (log$F_3$) and variability index (${\rm VI}$))
with the better  test results (e.g, $D > 0.2$ in KS test, or $p_3 < 1.00E-18$) 
{are selected in this work.}
Here, some of these 8 parameters  (e.g., spectral index and variability index) are also used in other recent works (e.g., \citealt{2014ApJ...782...41D},  \citealt{2016MNRAS.462.3180C} and \citealt{2017A&A...602A..86L}). 
{However, their research focus (e.g., aims and/or selected parameters and/or methods) are different from that of our work.
For instance, \cite{2014ApJ...782...41D} focused and identified ``AGN'' or ``non-AGN'' from 576 unassociated sources of the 2FGL catalogue using a neural network and a random forest SML algorithms; 
 \cite{2016MNRAS.462.3180C} focused and identified BL Lacs and FSRQs among the BCUs in the 3FGL catalogue using a neural network SML algorithm; 
The aim of \cite{2017A&A...602A..86L}  was, firstly, focused in identifying blazar candidates from the 3FGL unassociated sources,
second, to evaluate the BL Lacs or FSRQs from the blazar candidates (determined in their work and the BCUs that are already reported in the 3FGL catalogue) using multivariate classifications;
However, our research aim is to identify BL Lacs and FSRQs from the high-confidence clean sample of the 3LAC (3LAC Clean Sample) using 4 different SML algorithms (\emph{DT, RF, SVM and Mclust}).}
All the available observational data of the 8 parameters are directly obtained from the 3LAC Website version\footnote{http://www.asdc.asi.it/fermi3lac/} 
and LAT 4-year Point Source Catalog\footnote{https://heasarc.gsfc.nasa.gov/W3Browse/fermi/fermilpsc.html}.
However,  excluding 2 sources have no radio data and 2 missing data of curve significance ($C_S$), 
1416 sources (413 FSRQs, 603 BL Lacs  and 400 BCUs) are compiled in this work,  where, 400 BCUs are listed in Table \ref{DA_result}.

\begin{figure*}
\centering
    \includegraphics[height=18cm,width=18cm]{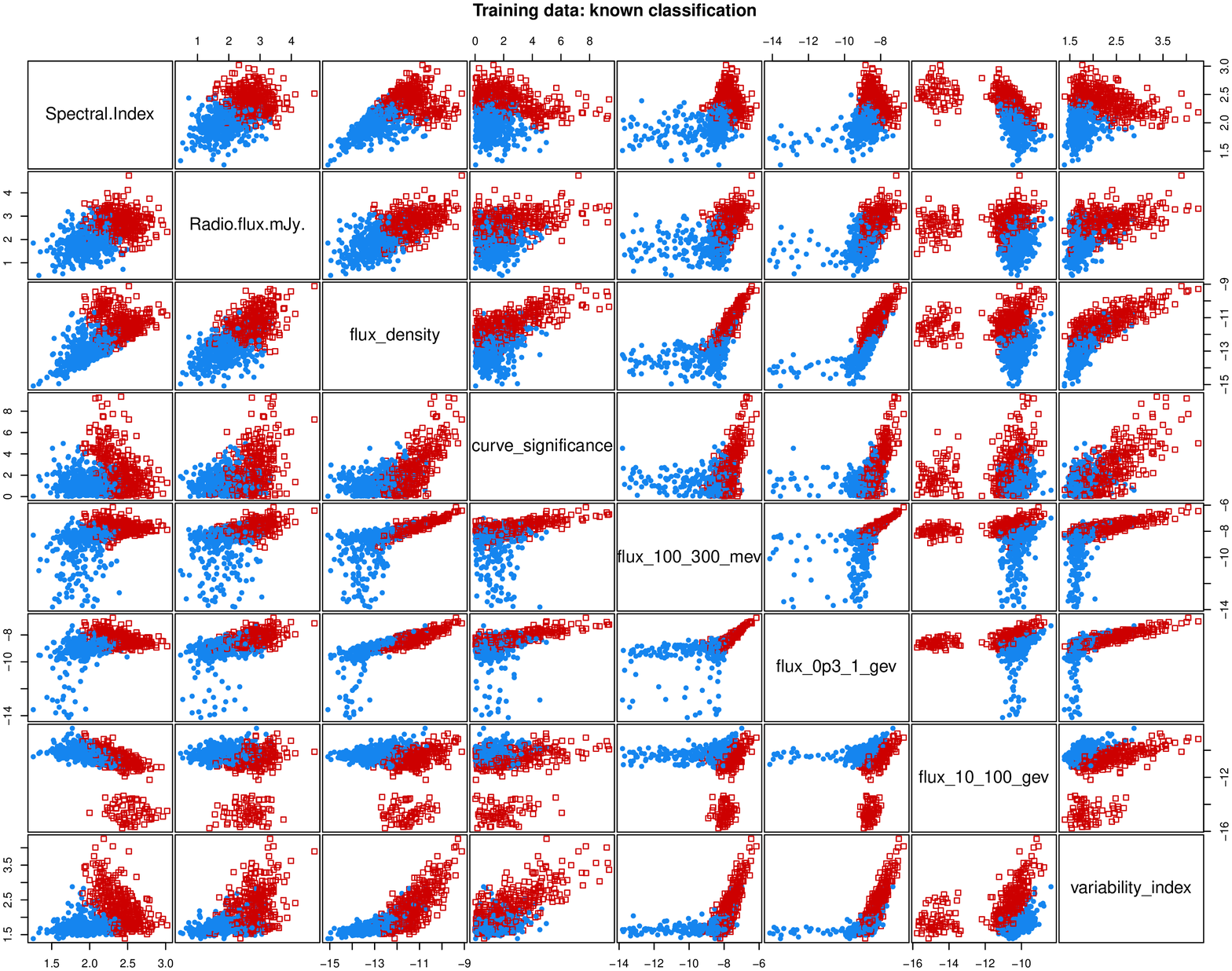}
  \caption{Classification plots for the $\gamma$-ray photon spectral index ($\Gamma_{\rm ph}$), radio flux  (log${\rm F_R}$), flux density (log$F_D$), curve significance ($C_S$), the integral  photon flux in 100 to 300 MeV (log$F_1$),  0.3 to 1 GeV (log$F_2$), 10 to 100 GeV (log$F_3$) and variability index (${\rm VI}$) 
  according to the known classification from $MclustDA$ for 4/5 of FSRQs (red empty squares) and BL Lacs (blue points) - training dataset respectively. }
 \label{fig_training}
\end{figure*}

\begin{figure*}
\centering
    \includegraphics[height=17cm,width=16cm,angle=270]{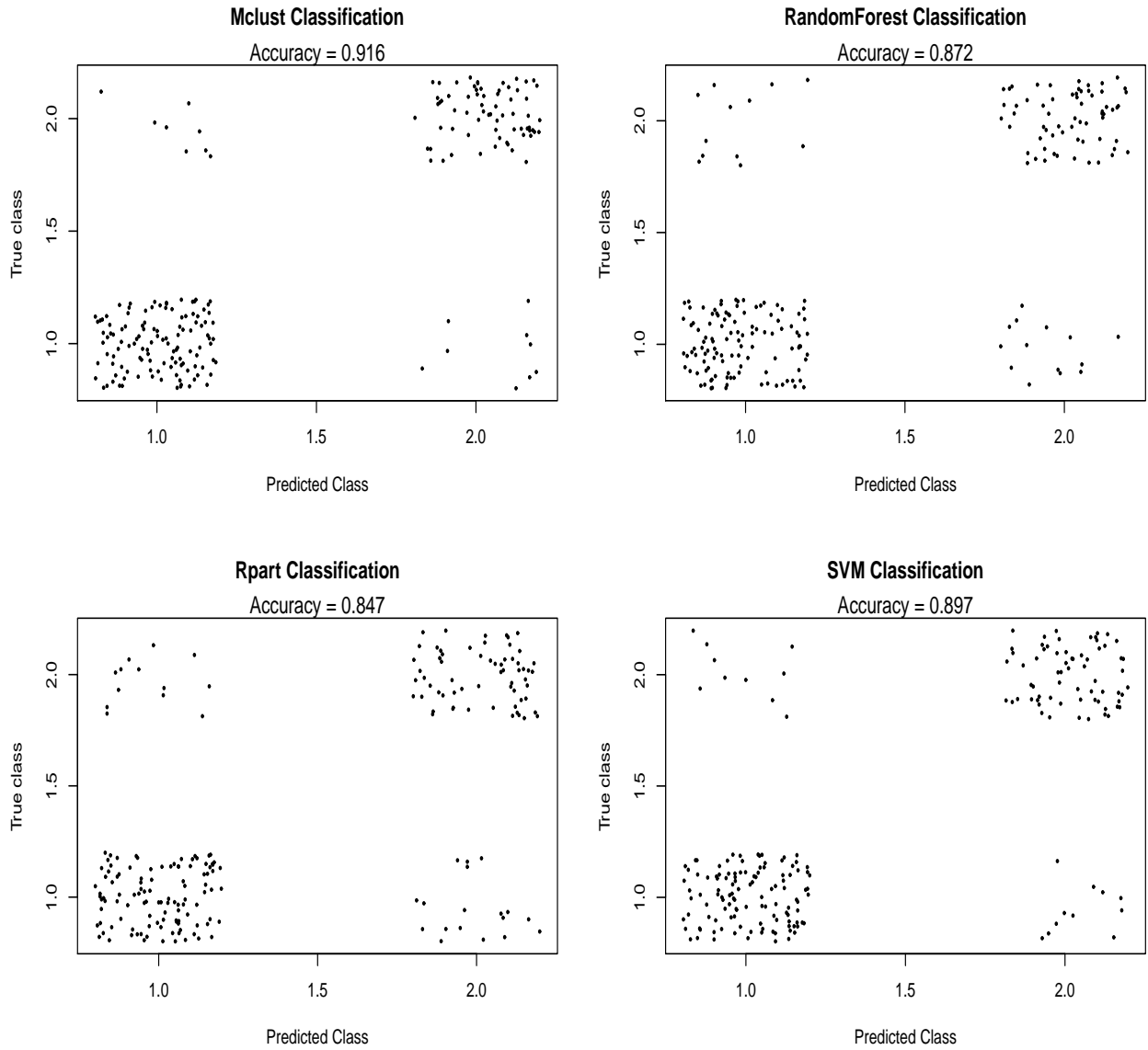}
  \caption{Misclassifications from four supervised classifiers 
  (supervised classifications of the training dataset: $Mclust$, Regression Trees, Random Forest and Support Vector Machines) 
  applied to the test set (1/5 of  604 BL Lacs  and 414 FSRQs).
 }
 \label{sub:fig1}
\end{figure*}

\begin{figure*}
\centering
    \includegraphics[height=18cm,width=18cm]{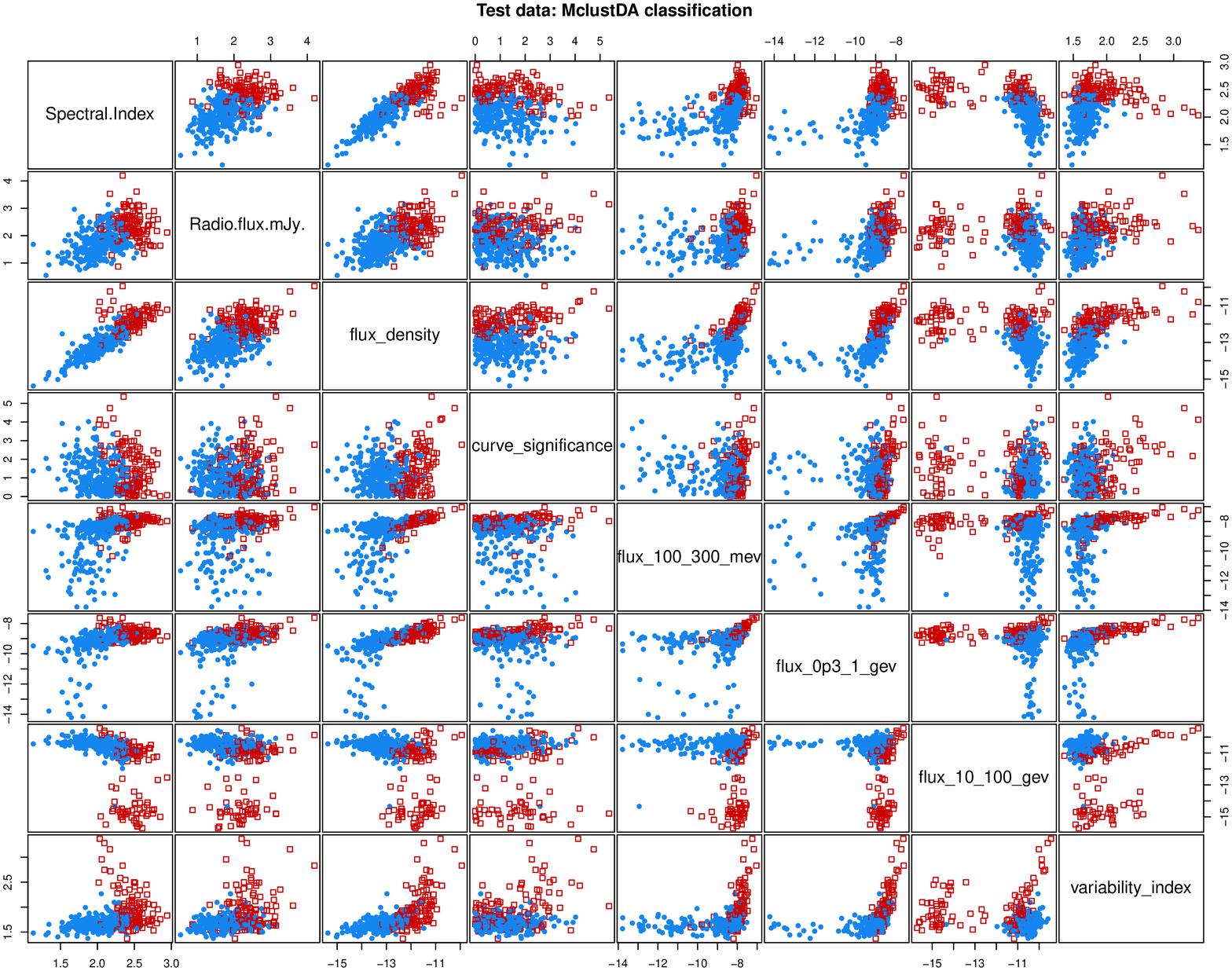}
  \caption{Classification plots for the $\gamma$-ray photon spectral index ($\Gamma_{\rm ph}$), radio flux  (log${\rm F_R}$), flux density (log$F_D$), curve significance ($C_S$),
the integral  photon flux in 100 to 300 MeV (log$F_1$),  0.3 to 1 GeV (log$F_2$), 10 to 100 GeV (log$F_3$) and variability index (${\rm VI}$)  
  according to the estimated classification from $MclustDA$ for fermi BCUs dataset (predicted dataset), 
  where red empty squares represent FSRQs and blue points represent BL Lacs identified using the function ``$predict.MclustDA()$".
 }
 \label{fig_test}
\end{figure*}

\section{Method} \label{sec:method}

{The fields of unsupervised and supervised machine learning}
{ provide many classification methods for predicting categorical outcomes,}
including \emph{logistic regression, decision trees, random forests, support vector machines, neural networks, 
Bayesian networks, Gaussian finite mixture models},
{and many others (e.g., see \citealt{feigelson_babu_2012,Kabacoff2015R} for the reviews).}

{In supervised learning}
(e.g., see \citealt{feigelson_babu_2012,Kabacoff2015R} for more detail),
{a dataset containing values for both the predictor variables and the outcome is divided into a training sample and a validation sample. Then one uses the training sample to develop a predictive model, while uses the validation sample to verify the accuracy. This dividing of data is essential for creating an effective model, since one needs a separate validation sample to make a realistic estimation of the effectiveness of the classification schemes developed on a training sample. Once an effective predictive model is created, one can use it to predict outcomes when only the predictor variables are known}
(e.g., see \citealt{feigelson_babu_2012,Kabacoff2015R} for more detail).

{In this section, a} brief introduction to \emph{DT, RF, SVMs and Mclust} 
{is provided.}
{``Decision trees"} 
{ aim to build a tree that can be used to classify new observations into one of two groups, by creating a set of binary splits on the predictor variables. 
They are popular in data-mining techniques (\citealt{Utgoff1989,Duda2001Pattern}), but very often, 
they tend to produce a large tree and suffer from overfitting}
(e.g., see \citealt{Leo1984Classification,Duda2001Pattern}).

{``Random forests"} 
{involve a large number of decision trees from a single training sample. 
The strategy is to enhance the classification by conducting votes among those many trees.}
The method is presented by {\cite{Breiman2001}}
and is applied to an astronomical dataset by {\cite{Breiman2003Random}}.
{It is a highly effective method, producing classification with better accuracy compared with other classification methods}
(e.g., \emph{Decision trees}) (e.g.,  see \citealt{JMLRv15delgado2014a}). 
{Additionally, it can handle problems with many observations and variables, and it can handle the cases where there are large amounts of missing data in the training set and where the number of variables is much greater than the number of observations. Another advantage of RF is that it produces OOB (out-of-bag) error rates and measures of variable importance. On the other hand, due to the large number of trees (default 500 trees), it is difficult to understand the classification rules and make communications.}

{$Support~vector~machines$} are a group of SML models that can be used for classification and regression {(\citealt{Vapnik1995,Vapnik2000})}.
{The mathematical theory behind it is to find the optimal hyperplane or a set of hyperplanes for separating classes in a high-dimensional space. This method produces accurate predicting models and is popular at present.}

{$Mclust$ (\citealt{mclust5})}
{is a powerful R package for model-based clustering, classification, density estimation to discriminant analysis. It is based on finite Gaussian mixture modelling and provides several tools for finite mixture models, including functions for parameter estimation using the EM algorithm}
(e.g., \citealt{FraleyandRaftery2002} and \citealt{mclust5}).

In order to estimate which approach is most accurate, and therefore to choose the best predictive solution, we define the quantity accuracy in binary classifications context. A function $performance()$ for calculating these statistics {\citep{Kabacoff2015R}} is provided. The $performance()$ function takes a table containing the true outcome (rows) and predicted outcome (columns) and returns the five accuracy measures. 
First, the number of 
true positives (a true positive is an outcome where the model correctly predicts the positive class),
true negatives (a true negative is an outcome where the model correctly predicts the negative class),
false positives  (a false positive is an observation with a positive classification is correctly identified as positive) and
false negatives (a false negative is an observation with a negative classification is correctly identified as negative)
are extracted. 
Next, these counts are used to calculate the sensitivity, specificity, positive and negative predictive values, and the accuracy {\citep{Kabacoff2015R}}.

In this work, 
{we}
will use {the $rpart$ packages in R} to create decision trees; 
the $randomForest$ package to fit random forests; 
and the e1071 package to build support vector machines. 
the $Mclust$ packages to fit Gaussian Mixture Modelling with the $MclustDA()$ function 
in the base R\footnote{http://www.r-project.org} installation.

\begin{table*}[ht]
	\centering
\caption{The predict and test results from four supervised classifiers to the BCUs and the test set.}
\begin{tabular}{c|lccccccc}
                \hline \hline
         	&    classifier    &	bll 	&	  fsrq  	&	Sensitivity	&	 Specificity&	 Positive Predictive	&Negative Predictive 	&	  Accuracy 	\\
	        \hline
(4/5)            &Mclust                &295	&     105    	&	0.895	&	0.929 	&	0.883 	&	0.937	&	0.916	\\                   
8                 &Random Forest&	308	&	92     	&	0.842	&	0.890	&	0.821	&	0.904	&	0.872	\\
parameters &rpart              &	279	&	121   	&	0.829	&	0.858	&	0.778	&	0.893	&	0.847	\\
seed=123    &svm           	&	301	&	99     	&	0.868 	&	0.913 	&	0.857 	&	0.921	&	0.897	\\
	           \cline{2-9}
                   &Mclust(EDDA)   &	273	&	127    	&	0.921	&	0.874	&	0.814	&	0.949	&	0.892	\\ 
                   &Forest(10000) &	309	&	91     	&	0.855	&	0.890	&	0.823	&	0.911	&	0.877	\\                 
                   &rpart (no pruned)  &293&	107   	&	0.776 	&	0.882	&	0.797 	&	0.868 	&	0.842	\\
                   &svm  (cost)         	&299	&	101   	&	0.829	&	0.921	&	0.863	&	0.900	&	0.887	\\                                              
	        \hline
(2/3)            &Mclust       	&	304	&	96     	&	0.845	&	0.931	&	0.891	&	0.900 	&	0.897	\\
8                  &Random Forest&	310	&	90     	&	0.837	&	0.897	&	0.843 	&	0.893	&	0.873	\\
parameters &rpart          	&	305	&	95     	&	0.770 	&	0.892 	&	0.825 	&	0.854 	&	0.844	\\
seed=123    &svm           	&	301	&	99     	&	0.859 	&	0.911	&	0.866	&	0.907	&	0.891	\\
	        \hline
(2/3)            &Mclust       	&	297	&	103     	&	0.819	&	0.896	&	0.843 	&	0.878 	&	0.864 	\\
8                  &Random Forest&	306	&	94     	&	0.833 	&	0.896	&	0.846	&	0.887	&	0.870 	\\
parameters &rpart          	&	300	&	100     	&	0.746       &	0.891	&	0.824	&	0.836 	&	0.832	\\
{\bf seed=321} &svm         &	299	&	101    	&	0.877	&	0.900 	&	0.858 	&	0.914	&	0.891	\\
	        \hline
(2/3)            &Mclust       	&	287	&	113     	&	0.864 	&	0.859 	&	0.812 	&	0.900	&	0.861 	\\
8                  &Random Forest&	302	&	98     	&	0.893 	&	0.884 	&	0.845	&	0.921 	&	0.888	\\
parameters &rpart          	&	291	&	109     	&	0.871 	&	0.854 	&	0.808	&	0.904 	&	0.861 	\\
{\bf seed=1234} &svm         &	292	&	108     	&	0.864 	&	0.864 	&	0.818	&	0.900 	&	0.864	\\	        
	        \hline	        
(4/5)             &Mclust       	&	289	&	111    	&	0.895	&	0.913	&	0.861	&	0.935	&	0.906	\\
4                 &Random Forest&	312	&	88     	&	0.776	&	0.882	&	0.797	&	0.868	&	0.842	\\
parameters &rpart          	&	305	&	95     	&	0.803	&	0.858	&	0.772	&	0.879	&	0.837	\\
seed=123     &svm           	&	305	&	95     	&	0.868	&	0.898	&	0.835	&	0.919	&	0.887	\\
	        \hline
(4/5)            &Mclust       	&	292	&	108    	&	0.908	&	0.858	&	0.793	&	0.940	&	0.877	\\
3                 &Random Forest&	304	&	96     	&	0.737	&	0.890	&	0.800	&	0.850	&	0.833	\\
parameters &rpart          	&	280	&	120   	&	0.842	&	0.843	&	0.762	&	0.899	&	0.842	\\
seed=123    &svm           	&	301	&	99     	&	0.855	&	0.890	&	0.823	&	0.911	&	0.877	\\
                 \hline  	        
\end{tabular}
\tablecomments{
Column 1 shows the different combinations of parameters (8, 4 or 3 parameters, see Section 5), different quality (e.g., seed=123, 321 or 1234) and different quantity (e.g., 4/5 or 2/3) of the training samples.
Column 2 gives the classifiers; Column 3 and Column 4 give the number of BL Lacs and FSRQs predicted by a supervised classifier (using supervised machine-learning techniques) to the BCUs (predicted dataset).
A classifier's sensitivity, specificity, positive predictive power, and negative predictive power are shown in Column 5, Column 6,  Column 7 and  Column 8 respectively;
The accuracy of a classifier  is reported in Column 9.
 }
 \label{tab_result}
\end{table*}

\section{Results} \label{sec:result}

The selected sample (1416 sources) includes 1016 sources (413 FSRQs and 603 BL Lacs) with known optical classification 
 and 400 BCUs  with unidentified the optical classification in $Fermi$ 3FGL catalog.
 In supervised learning, 
 we randomly ({random seed =123}) assign approximately 4/5 of the observations (known classification: 603 BL Lacs and 413 FSRQs) to the training dataset, and the remaining ones to the validation dataset (test set) in the 8-dimensional parameter space as described  in Section 2.
 {All the 400 BCUs are viewed as  a sample for prediction (as forecast dataset).}
The training set has 813 blazars (476 BL Lacs, 337 FSRQs), and the validation set has 203 blazars (127 BL Lacs, 76 FSRQs).
The training dataset is used to create classification schemes using \emph{a decision tree, a random forest, a support vector machine and a Gaussian Mixture Modelling ($Mclust$)}. 
Where, in order to simplify calculating, all the default settings for each of 4 classification function (e.g., $MclustDA()$, $randomForest()$, $rpart()$ and $svm()$ function) are used in this work. 
The validation dataset is used to evaluate the effectiveness of these schemes.
Using an effective predictive model that is developed using the data in the training set to forecast dataset, one can predict outcomes (a BCU belongs to BL Lacs or FSRQs) in situations where only the predictor variables are known. Here, the main R steps can be obtained from a public website\footnote{https://github.com/ksj7924/Kang2019ApJRcode}.

In $Mclust$ discriminat analysis, 
{we use the function} $MclustDA()$ 
with \emph{MclustDA}  model (modelType = ``MclustDA", where each known classification is modeled by a finite mixture of Gaussian distributions with a number of components and covariance matrix structures being different between classes, named as MclustDA , see, e.g., \citealt{FraleyandRaftery2002} and \citealt{mclust5})
to the training dataset{. The largest} BIC (Bayesian Information Criterion) value of {-6370.124} was obtained using the VEV model (assuming clusters having ellipsoidal distributions described by variable volumes, equal shapes and variable orientations) with a 4-component mixture distribution for 476 BL Lacs; 
and EVE model (assuming clusters having ellipsoidal distributions described by equal volumes, variable shapes and  equal orientations) with a 4-component mixture distribution for 337 FSRQs; 
based on the training sample. The training error rate $\simeq$ {0.116} is also obtained based on the the function $MclustDA()$. 

In order to test the result of $Mclust$ supervised learning ($MclustDA()$ discriminate analysis), the $predict.MclustDA()$ function are used to the test set,  so the test error rate $\simeq$ {0.084} is reported. Here, we also compute the classification error using cross-validation. A cross-validation error  $\simeq$  {0.138} (which is approximately consistent with the training error rate $\simeq$ {0.116}) can be computed using the cvMclustDA() function, which by default use nfold = 10 for a 10-fold cross-validation. The classification for the training {dataset} from $MclustDA$ are shown in Figure \ref{fig_training}. 
In order to evaluate the utility of a classification scheme, the $performance()$ function is performed and  returns
Sensitivity = {0.895},
Specificity = {0.929},
Positive Predictive Value = {0.883},
Negative Predictive Value = {0.937},
Accuracy = {0.916}
(see Table \ref{tab_result} and Figure \ref{sub:fig1}).
Using the function ``$predict.MclustDA()$" for classifying predicted dataset (BCUs), we obtain 295 BL Lacs and 105 FSRQs  (see Figure \ref{fig_test}, Table \ref{tab_result} and machine-readable supplementary material in Table \ref{DA_result}) from the 400 BCUs (2 sources have no radio data are excluded).

{In {$Random forests$} discriminat analysis,
using the $randomForest()$ function (the default number of trees is 500) in the random-Forest {R} package \citep{randomForest} to the training dataset, 
OOB (out-of-bag) estimate of  error rate = {0.124} {was} obtained.}
Where random forests also provide a natural measure of variable importance: the $Gini$ coefficient (see Table \ref{tab:test}) of 
Spectral Index = {90.89},
flux density  =     {74.98},
Radio flux =        {55.55},
variability index  =  {45.65},
flux 10$-$100 GeV   =      {47.94},
flux 100$-$300 MeV  =     {32.89},
flux 0.3$-$1 GeV =           {24.04},
curve significance   =        {22.20}{, which suggests} 
that Spectral Index is the most important variable and curve significance is the least important among the 8 selected parameters.
Applying the predictive model obtained from the random forest to the validation sample, 
the validation sample is classified and the predictive 
Sensitivity =  {0.842},
Specificity =  {0.890},
Positive Predictive Value =   {0.821},
Negative Predictive Value =  {0.904}
and
Accuracy =  {0.872},
are calculated (see Table \ref{tab_result} and Figure \ref{sub:fig1}).
{Applying} the random forest predictive model to the forecast dataset, 
{we} obtain  308 BL Lacs and 92 FSRQs  (see Table \ref{tab_result} and machine-readable supplementary material in Table \ref{DA_result}) from the 400 BCUs.

{For the training sample, a decision tree is grown using the $rpart()$ function in R package \citep{rpart}.}
However, unfortunately,  the tree sometimes becomes too large and suffers from overfitting {(e.g., \citealt{Leo1984Classification,Duda2001Pattern})}.
To make up for the deficiency, a $prune()$ function is used to prune back the tree in the $rpart$ package.
And then a tree with the desired size can be obtained.  
Using it to the validation sample, the 
Sensitivity =  {0.829},
Specificity =  {0.858},
Positive Predictive Value =  {0.778},
Negative Predictive Value =  {0.893},
Accuracy =  {0.847}, are shown (see Table \ref{tab_result} and Figure \ref{sub:fig1}).
Then using it to the forecast dataset, 279 BL Lacs and 121 FSRQs  are obtained 
(see Table \ref{tab_result} and machine-readable supplementary material in Table \ref{DA_result}).

Finally, Support vector machines is also 
{applied} to the training sample.
{The $svm()$ function in the e1071 R package \citep{e1071} is used.}
Using the optimal predictive model obtained from SVMs to the validation sample, the
Sensitivity =  {0.829},
Specificity =  {0.921},
Positive Predictive Value =  {0.863},
Negative Predictive Value =  {0.900},
Accuracy =  {0.887}, are printed
(see Table \ref{tab_result} and Figure \ref{sub:fig1}).
Also using the optimal predictive $SVMs$ model to the forecast dataset, 301 BL Lacs and 99 FSRQs  are obtained 
(see Table \ref{tab_result} and machine-readable supplementary material in Table \ref{DA_result}).

\begin{table*}[ht]
	\centering
\caption{The comparison results for four supervised classifiers and other resent works}
\begin{tabular}{c|lccccccch}
                \hline \hline
            &  ${\rm Mclust}$ & ${\rm DT}$  &  ${\rm RS}$  &	${\rm SVM}$  &  $Y$ &	{$M$} 	&{$LP$}  	&	  Chi16 	\\
	        \hline
 $-$           &                       &      	&               	&	         	&	          	&	244    	&	3       	&	 	\\                   
bll              &295                 &	253	&	277   	&	274   	&	218   	&	{\bf 47}     	&	228    	&	244	\\
fsrq$^a$     &0                     &	42	&	18     	&	21     	&	76     	&	{\bf 4}       	&	36      	&	22	\\
unc            &              	 &		&	         	&	         	&	 1      	&	          	&	28     	&	29	\\
rate           &             &$\sim$14.2\%  &$\sim$6.1\%  	&$\sim$7.7\%  	&$\sim$25.9\% &$\sim$7.8\%  	&$\sim$13.6\%  &$\sim$8.3\%  \\
	        \cline{1-9}
 $-$           &                       &      	&               	&	         	&	         	&	96     	&	          	&	 	\\                   
bll$^b$        &0                    &	26	&	31     	&	27     	&	8       	&	2       	&	14     	&	29	\\
fsrq           &105                &	79	&	74     	&	78     	&	97     	&	7       	&	83      	&	54	\\
unc           &                      &		&	         	&	         	&	         	&	          	&	8       	&	22	\\
rate           &             &$\sim$24.8\%  &$\sim$29.5\% &$\sim$25.7\%  &$\sim$7.6\% &$\sim$22.2\%  	&$\sim$14.4\%   &$\sim$34.9\%\\
                \hline                                              
 $-$           &                       &      	&               	&	         	&	          	&	197   	&	3       	&	 	\\                   
bll              &\multicolumn{4}{c}{{246}}                 	                     	&	207   	&	46     	&	213    	&	224	\\
                 \cline{2-5} 
fsrq$^a$      &\multicolumn{4}{c}{{0}}                 	                          	&	38     	&	3       	&	8       	&	5	\\
unc            &              	 &		&	         	&	         	&	 1      	&	          	&	22     	&	17	\\
rate           &               	 &		&	         	&	         	 &$\sim$15.5\% &$\sim$6.1\%  	&$\sim$3.6\%   &$\sim$2.2\%\\
	        \cline{1-9}
 $-$           &                       &      	&               	&	         	&	         	&	57     	&	          	&	 	\\                   
bll$^b$         &\multicolumn{4}{c}{{0}}                 	                          	&	0       	&	0       	&	0       	&	6	\\
                      \cline{2-5} 
fsrq               &\multicolumn{4}{c}{{64}}                 	                         &	64     	&	7       	&	63     	&	46	\\

unc           &                      &		&	         	&	         	&	         	&	          	&	1       	&	12	\\
rate           &               	 &		&	         	&	         	 &$\sim$0\%     &$\sim$0\%  	&$\sim$0\%      &$\sim$11.5\%\\
                \hline                                              
\end{tabular}
\tablecomments{
Column 1 shows the different classification ($-$ represents the number of mismatch by cross comparison,  ``{bll}" ,``{fsrq}" and  ``{unc}" indicate  BL Lac,  FSRQ and uncertain type respectively, 
where $^a$ and $^b$  indicate the number of the disagreement (misjudged as FSRQs or BL Lacs)), and the mismatch rate (e.g., rate = 4/(4+47)\% $\sim$ 7.8\% ).
The comparison results of  \emph{$Mclust$ Gaussian Mixture Modelling (Mclust), decision tree (DT), random forest (RS), and support vector machine (SVM)} are listed in Column 2, Column 3, Column 4, and Column 5.
The results of cross comparison with \citealt{2017ApJ...838...34Y} (Y); \citealt{2016Ap&SS.361..337M} (M); \citealt{2017A&A...602A..86L} (LP) and \citealt{2016MNRAS.462.3180C} (Chi16)
 are shown in Column 6,  Column 7,  Column 8 and  Column 9 respectively.
 }
 \label{tab_disscusion}
\end{table*}

\section{Discussions and Conclusions}\label{sec_conclusions}

In this work, 
one
try to evaluate the potential classification of $Fermi$ BCUs using the supervised machine learning (discriminant  analysis).
We use 4 methods (\emph{DT, RF, SVMs and Mclust}) to perform the discriminant  analysis for 8 parameters ($\Gamma_{\rm ph}$, log${\rm F_R}$, log$F_D$, $C_S$, log$F_1$,  log$F_2$, log$F_3$ and log ${\rm VI}$).
All the 4 classifiers
{perform exceedingly well and produce accurate and effective  forecast of the classification of $Fermi$ BCUs. Compared with the results of these methods, $Mclust$ Gaussian Mixture Modelling is the most promising (see Table \ref{tab_result}) for our training sample (4/5, seed=123).  }

{
FSRQs have stronger emission lines (${\rm EW\geq5\AA}$), while the BL Lacs have weak (${\rm EW < 5\AA}$) or no emission lines (e.g., \citealt{1995PASP..107..803U});
FSRQs show higher luminosity than that of BL Lacs (e.g., see \citealt{1998MNRAS.299..433F}; \citealt{2011MNRAS.414.2674G} and \citealt{2017MNRAS.469..255G}); 
Based on the 3LAC catalogue, \cite{2015ApJ...810...14A} argued that FSRQs tend to have softer spectra,  stronger variability and lower peak frequencies in both synchrotron and IC components than BL Lacs; 
And many others.
These distinctions suggest different physical origin between in FSRQs and in BL Lacs (e.g., \citealt{2016RAA....16...54B,2016RAA....16..173F,2018SCPMA..61e9511Y,2019MNRAS.482L..80B}).
The synchrotron radiation peak frequency of FSRQs is significantly lower than that of BL Lacs (e.g., \citealt{1998MNRAS.299..433F}; \citealt{2009ApJ...700..597A,2010ApJ...715..429A,2011ApJ...743..171A,2015ApJ...810...14A,2017MNRAS.469..255G}), this imply that more electron populations lose their energy through synchrotron cooling in FSRQs. In this scenario, we could expect stronger radio emission in FSRQs. 
For gamma-ray band, it is commonly believed that the gamma-ray radiation in BL Lacs originate from a pure  synchrotron self-Compton (SSC) process, 
(e.g., \citealt{1997A&A...320...19M}; \citealt{2004ApJ...601..151K}; \citealt{2011ApJ...728..105Z}; \citealt{2014ApJ...788..104Z}; \citealt{2014MNRAS.442.3166Z}; \citealt{2017ApJ...842..129C}; \citealt{2018MNRAS.478.3855Z}),
while that in FSRQs come from  SSC+EC (external Compton) processes
(e.g., \citealt{1999ApJ...515..140S}; \citealt{2002ApJ...581..127B}; \citealt{2011ApJ...735..108C}; \citealt{2014ApJS..215....5K}; \citealt{2016MNRAS.461.1862K}; \citealt{2016MNRAS.457.3535Z}; \citealt{2017ApJS..228....1Z}).
This indicates, for FSRQs, a complex physical process can be expected in Fermi energy bands. 
The Fermi energy spectrum in FSRQs could be resulted from the spectrum that is superposed other spectra components 
(e.g., \citealt{2013ApJ...764..113Z}; \citealt{2016A&A...585A...8Z}; \citealt{2017ApJ...837...38K}).
The Fermi  band of FSRQs locating at intersection of both synchrotron self-Compton component and external Compton component could result to a  more complex observational features
(e.g.,  \citealt{2009ApJ...700..597A,2010ApJ...715..429A}; \citealt{2011ApJ...743..171A,2015ApJ...810...14A}).
Other physical origins (e.g., mass accretion rate on to the central black hole) are also discussed in resent works (e.g., \citealt{2019MNRAS.482L..80B}).
The more fundamental physical origins between in FSRQs and in BL Lacs require further discussion in the future.
}

One checks the results for different combinations of parameters.
Based on the $Gini$ coefficient (a natural measure of variable importance) in random forests supervised learning (see Column 9  in Table \ref{tab:test}),
one selects part parameters with a higher $Gini$ coefficient 
(4 parameters: $\Gamma_{\rm ph}$, log${\rm F_R}$, log${\rm VI}$ and log$F_D$; 
or 3 parameters: $\Gamma_{\rm ph}$, log${\rm F_R}$,  log$F_D$)
to discriminant  analysis also using the 4 methods.
We find that the predictive accuracy will be smaller than that of 8 parameters.
However, $Mclust$ Gaussian Mixture Modelling also tends to be more accurate compared with other classification methods for the different testing variables in combination (see Table \ref{tab_result}, e.g., 8 parameters, 4 parameters and 3 parameters) for our training sample. 
In general, which implies more parameters and more accuracy, but it is unstable for the different classification methods (see Table \ref{tab_result} for the details).

However, we should note that the predictive accuracy and results may be affected by the training dataset and validation dataset.
When one randomly ({seed=123}) assigns approximately 2/3 of the known classification blazars (603 BL Lacs and 413 FSRQs) to the training dataset (677 blazars: 399 BL Lacs and 278 FSRQs)
 and the remaining ones to the validation dataset (339 blazars: 204 BL Lacs and 135 FSRQs)
 in the 8-dimensional parameter space as in Section 4.
The predictive accuracy (Accuracy =  {0.897}) and results (304 BL Lacs and 96 FSRQs predicted from the 400 BCUs)  
are slightly different with that of the ({4/5, seed=123}) training and validation samples (see Table \ref{tab_result}) in $Mclust$ discriminant  analysis.
And other methods also show similar results.
Also, for the randomly samples (e.g., randomly seed =123, {=321, or =1234}),  the results and accuracy are also different (see Table \ref{tab_result}),
where the most accuracy are obtained in the support vector machines (seed=321) or in the Random forests (seed=1234) respectively.
These suggest the results of discriminant analysis (supervised learning) are significantly affected by the quality (e.g., seed=123, 321 or 1234) and quantity 
(e.g., changed from 813 blazars (4/5, see Section 4) to  677 blazars (2/3)) of the training samples.
Where, sometimes, the support vector machines or the Random forests  yield a higher accuracy, which is consistent with other works (e.g.,  \citealt{JMLRv15delgado2014a}).

In addition, we also should note that all the default settings for each of the 4 classification function
 (e.g., $MclustDA()$, $randomForest()$, $rpart()$ and $svm()$ function) are used in Section 4.
For each different classification method, choosing of calculation model and setting of each parameter in fitting function 
(e.g., the ``modelType = {MclustDA or EDDA (Eigenvalue Decomposition Discriminant Analysis, e.g., see \citealt{mclust5})})" in $MclustDA()$ function , the ``tree = {500 or 10000}"  in $randomForest()$ function, if a $prune()$ function  is used in $rpart()$,
``gamma={0.1 or 0.01}'' and ``cost={1 or 1000}''  in $svm()$ function) 
can also affect predictive models, accuracy and results (see the labels ``EDDA", ``Forest", ``no pruned" and ``cost" in Table \ref{tab_result}).
About how to select the appropriate parameter settings are need to further address in future, which also is beyond the scope of this work.

{
We compare the results of $Mclust$  SML algorithms with that of  other three SML algorithms.
We find that, for BL Lacs, 
 253, 277 and 274 (mean about 91\%) BL Lac candidates in  \emph{DT, RF and SVMs} match with $Mclust$ (295) BL Lacs candidate sample (see Table \ref{tab_disscusion}); 
unfortunately, 42, 18 and 21 (mean about 9\%) sources classed as FSRQ do not match the $Mclust$ (295) BL Lacs candidates respectively.
For FSRQs, 79, 74 and 78  (mean about 73\%)  FSRQ candidates match the results (105 FSRQ candidates) of $Mclust$ method; but 26, 31 and 27 (mean about 27\%) sources do not match the subset of $Mclust$ 105 FSRQ candidates respectively (see Table \ref{tab_disscusion}). 

We also compare the results of $Mclust$ algorithms with other resent similar results 
(e.g., \citealt{2016MNRAS.462.3180C}; \citealt{2016Ap&SS.361..337M}; \citealt{2017A&A...602A..86L}; \citealt{2017ApJ...838...34Y}). 
After cross comparison with the results of  \cite{2017A&A...602A..86L} using multivariate classifications,
in the subset of 295 BL Lac candidates (see $Mclust$ method), we find that 3 sources do not match sources and 28 sources did not provide a clear classification in \cite{2017A&A...602A..86L}.
$Mclust$ prediction is in accordance with \cite{2017A&A...602A..86L} for 228 objects (about 77\%) and  is inconsistent for 36 (about 12\%). 
For the subset of 105 FSRQ candidates, 83 objects are in agreement with and 14 objects are in disagreement with $Mclust$ prediction, and 8 objects do not provide a clear classification \citep{2017A&A...602A..86L}.
Also, in the subset of 295 BL Lac candidates, $Mclust$ prediction is in accordance with \cite{2016MNRAS.462.3180C} 
using artificial neural networks (ANN) machine-learning techniques for 244 sources and is inconsistent for 22, 29 objects do not provide a clear classification  \citep{2016MNRAS.462.3180C}.
For the subset of 105 FSRQ candidates, only 54 objects are in agreement with and 29 objects are in disagreement with $Mclust$ prediction, and 22 objects do not provide a clear classification  \citep{2016MNRAS.462.3180C}.
For comparison with \cite{2017ApJ...838...34Y} performed a statistical analysis of the broadband spectral properties (e.g., spectral indices in the gamma-ray, X-ray, optical, and radio bands) of blazars,  
the similar results are also shown.  218 BL Lac candidates and 97 FSRQ candidates are in agreement with $Mclust$ prediction (295 BL Lac and 105 FSRQ candidates);
4 BL Lac or 8 FSRQ candidates objects are in disagreement with $Mclust$ prediction (see Table \ref{tab_disscusion}).
Most of the results of $Mclust$ SML are consistent with that  (47  BL Lacs $\sim$92.2\% and 7 FSRQs $\sim$77.8\%) of  \cite{2016Ap&SS.361..337M} using optical spectroscopic observations.
}{
However, a fraction of sources (4 BL Lacs and 2 FSRQs) are misjudged using $Mclust$ SML.
These results suggest that it is a good overall agreement in these SMLs and other resent results.
However,  SML  algorithms probably lead to some misjudgments in evaluating the potential (optical) classification of blazars. 
Only the  optical spectroscopic observations  is still most efficient and accurate way to determine the real nature of these sources.
}

{
When we combine the results of these 4 methods, 246 BL Lacs and 64 FSRQs candidates  are obtained (see Table \ref{tab_disscusion}). 
Although the quantity has been decreased, the quality has been improved.
The mismatch rate (e.g., rate = 4/(4+47)\% $\sim$ 7.8\%, see Table \ref{tab_disscusion}) drops significantly from about 25.9\%, 7.8\%, 13.6\% and 8.3\% to about 11.5\%, 6.1\%, 3.6\% and 2.2\% for BL Lac, 
and from about 7.6\%, 22.2\%, 14.4\% and 34.9\% to about 0\%, 0\%, 0\% and 11.5\%  for FSRQs
in comparison with the results of \cite{2017ApJ...838...34Y},  \cite{2016Ap&SS.361..337M}, \cite{2017A&A...602A..86L}  and  \cite{2016MNRAS.462.3180C} respectively (see Table \ref{tab_disscusion}),
which suggest that the better results can be obtained by applying multiple methods simultaneously.
}

Although the discriminant  analysis can return the probabilities $P_{Bi}$ and $P_{Fi}$ that a BCU $i$ belongs to the BL Lacs (B) or FSRQs (F) classifications, respectively
 (e.g., in $Mclust$ method, see Table \ref{DA_result} and a machine-readable supplementary material).
However, it should be noted that
the error of the supervised machine learning is still very large (accuracy is still not high enough) in the work,
{where, the accuracy is less than 92\%}.
It also probably leads to some misjudgments that some FSRQs are falsely classified as BL Lacs, and vice versa  (see discussion above).
These result may be biased, or only be an apparent phenomenon, or hide the essential difference between BL Lacs and FSRQs (e.g., see \citealt{2018arXiv181206025B} for the reviews and \citealt{2019MNRAS.482L..80B}),  
which needs further consideration.
Although we did not conclusively evaluate their potential classifications (FSRQs and BL Lacs) using SML, 
it may be helpful for source selections in the spectroscopic observation campaigns in the future performing a spectroscopic 
and photometric campaign, further diagnosing  their optical classification of BCUs (e.g., see \citealt{2017ApJ...838...34Y,2013ApJS..207...16M} for some discussion),
or provide some clues for future studying of spectroscopic and photometric.

Finally, it must be highlighted that, in this work, 
our results are obtained only by supervised machine learning the data obtained from Fermi catalogue, 
adding no external data {(obtained from other archives)}, 
and have not done any of the fittings ourselves. 
The limited sample (made of mostly bright, firmly classified sources, excluded fainter sources) 
is used to diagnose the optical classifications of the BCUs in 3LAC Clean Sample. 
Selection effects of the direct observational data because of detection thresholds and energy bands in instruments 
may affect the source distributions and affect the results of the analysis in this work.

However, it should point out that {each} of these classifiers (\emph{DT, RF,  SVMs and Mclust}) performed exceedingly well on each of the accuracy measures.
The results of SML are in agreement with each classifiers and other resent results (\citealt{2016MNRAS.462.3180C}; \citealt{2016Ap&SS.361..337M}; \citealt{2017A&A...602A..86L}; \citealt{2017ApJ...838...34Y}),
which suggests that the SML can provide an effective and easy method to evaluate the potential classification of BCUs.
The evaluating results show the approach (SML) is valid and robust (see Table \ref{tab_result}).
It is about 1:3 ratio between FSRQs and BL Lacs predicted from the 400 BCUs for any SML algorithms.
 {Here, we also should note that 1:4 ratio (64 FSRQs  and 246 BL Lacs) was obtained by combinating the results of these 4 methods (see Table \ref{tab_disscusion}).
Whether the true ratio is 1:3 or 1:4 or others needs further verification.}
However, $Mclust$ Gaussian Mixture Modelling tend to be more accurate compared with other classification methods for the different testing the variables in combination 
(see Table \ref{tab_result}, e.g., 8 parameters, 4 parameters and 3 parameters) for our training sample. 
Although there are a number of factors influencing the accuracy of SML.
However, this work provides some simple methods to distinguish the BL Lacs or FSRQs with the probabilities $P_{Bi}$ and $P_{Fi}$ (see Table \ref{DA_result}) from BCUs 
based on the direct observational data. 
A more preferable statistical approach, that uses the a large and more complete sample (e.g., the upcoming 4LAC) are needed to further test and address the issue.

{\bf\color{blue}********************}

\section*{Acknowledgements}
We thank the anonymous referee for very constructive and helpful comments and suggestions, which greatly helped us to improve our paper,
and thanks for Lv xin's help in language and writing.
This work is  partially supported by the National Natural Science Foundation of China  
(Grant Nos.11763005, 11873043, 11847091, 11733001, 11622324, U1531245, and 11573009), 
the Science and Technology Foundation of Guizhou Province (QKHJC[2019]1290),
the Research Foundation for Scientific Elitists of the Department of Education of Guizhou Province (QJHKYZ[2018]068),
the Open Fund of Guizhou Provincial Key Laboratory of Radio Astronomy and Data Processing (KF201811), 
the Natural Science Foundation of the Department of Education of Guizhou Province (QJHKYZ[2015]455),
the Physical Electronic Key Discipline of Guizhou Province (ZDXK201535),
the Research Foundation for Advanced Talents of Liupanshui Normal University (LPSSYKYJJ201506),
the Research Foundation of Liupanshui Normal University (LPSSY201401),
the cultivation project of Master's degree of Liupanshui Normal University(LPSSYSSDPY201704)
the Key Disciplines Construction Project of Liupanshui Normal University (LPSZDZY201803), 
the Physics  Key Discipline of Liupanshui normal university (LPSSYZDXK201801),
and the Experimental Teaching Demonstration Center of Liupanshui Normal University (LPSSYsyjxsfzx201801).

\clearpage
\begin{longrotatetable}
\begin{deluxetable*}{lllllllllllllllllllll}
\tablecaption{The classification of $Fermi$ BCUs \label{DA_result}}
\tablewidth{700pt}
\tabletypesize{\scriptsize}
\tablehead{ 
\colhead{3FGL Name}      & \colhead{log$F_R$}             & \colhead{$\Gamma_{ph}$}     & \colhead{log$F_D$}                    & \colhead{$C_S$} 
& \colhead{log$F_1$}         & \colhead{log$F_2$}                & \colhead{log$F_3$}              & \colhead{log$VI$}                       & \colhead{$P_{Bi}$} 
& \colhead{$P_{Fi}$}          & \colhead{${\rm Class}_M$}    & \colhead{${\rm Class}_{DT}$}    & \colhead{${\rm Class}_{RS}$}     & \colhead{${\rm Class}_{SVM}$}   
& \colhead{$Y$}                 & \colhead{$M$}                        & \colhead{$LP$}  &   \colhead{$Chi16$} 
} 
\startdata
3FGL J0002.2$-$4152	&	1.121	& 2.089	&	-13.135	&	0.200	&	-8.275	&	-9.247	&	-10.587	&	1.751	&	100.00\%	&	0.00\%	&	 bll 	&	 bll 	&	 bll 	&	 bll 	&	bll	&	$-$	&	unc	&	bll   	\\
3FGL J0003.2$-$5246	&	1.815	& 1.895	&	-13.699	&	0.909	&	-8.182	&	-12.082	&	-10.526	&	1.656	&	90.56\%	&	9.44\%	&	 bll 	&	 bll 	&	 bll 	&	 bll 	&	bll	&	$-$	&	bll	&	bll   	\\
3FGL J0017.2$-$0643	&	1.973	& 2.116	&	-12.955	&	0.948	&	-9.233	&	-9.206	&	-10.979	&	1.573	&	99.88\%	&	0.12\%	&	 bll 	&	 bll 	&	 bll 	&	 bll 	&	bll	&	$-$	&	bll	&	bll   	\\
3FGL J0019.1$-$5645	&	1.782	& 2.391	&	-12.488	&	0.525	&	-8.003	&	-9.207	&	-10.513	&	1.798	&	99.86\%	&	0.14\%	&	 bll 	&	 bll 	&	 bll 	&	 bll 	&	fsrq	&	$-$	&	fsrq	&	bll   	\\
3FGL J0028.6$+$7507	&	1.909	& 2.342	&	-12.298	&	0.407	&	-7.975	&	-8.724	&	-11.472	&	1.577	&	0.93\%	&	99.07\%	&	 fsrq 	&	 bll 	&	 bll 	&	 bll 	&	fsrq	&	$-$	&	bll	&	bll   	\\
3FGL J0030.2$-$1646	&	0.979	& 1.647	&	-13.801	&	1.326	&	-12.006	&	-14.218	&	-10.320	&	1.808	&	100.00\%	&	0.00\%	&	 bll 	&	 bll 	&	 bll 	&	 bll 	&	bll	&	bll	&	bll	&	bll   	\\
3FGL J0030.7$-$0209	&	2.473	& 2.378	&	-11.547	&	3.137	&	-7.836	&	-8.406	&	-14.839	&	2.545	&	0.00\%	&	100.00\%	&	 fsrq 	&	 fsrq 	&	 fsrq 	&	 fsrq 	&	fsrq	&	$-$	&	fsrq	&	fsrq     	\\
3FGL J0031.3$+$0724	&	1.086	& 1.824	&	-13.917	&	1.060	&	-9.117	&	-10.830	&	-10.359	&	1.519	&	99.95\%	&	0.05\%	&	 bll 	&	 bll 	&	 bll 	&	 bll 	&	bll	&	$-$	&	bll	&	bll   	\\
3FGL J0039.0$-$2218	&	2.069	& 1.715	&	-14.096	&	2.687	&	-11.200	&	-9.461	&	-10.684	&	1.563	&	96.61\%	&	3.39\%	&	 bll 	&	 bll 	&	 bll 	&	 bll 	&	bll	&	$-$	&	bll	&	bll   	\\
3FGL J0039.1$+$4330	&	0.913	& 1.963	&	-13.352	&	1.853	&	-8.533	&	-9.463	&	-10.564	&	1.549	&	100.00\%	&	0.00\%	&	 bll 	&	 bll 	&	 bll 	&	 bll 	&	bll	&	$-$	&	bll	&	bll   	\\
3FGL J0040.3$+$4049	&	1.683	& 1.132	&	-15.375	&	1.377	&	-8.680	&	-9.480	&	-10.426	&	1.481	&	100.00\%	&	0.00\%	&	 bll 	&	 bll 	&	 bll 	&	 bll 	&	bll	&	$-$	&	bll	&	bll   	\\
3FGL J0040.5$-$2339	&	1.730	& 1.946	&	-13.676	&	1.383	&	-11.375	&	-9.226	&	-10.564	&	1.692	&	99.05\%	&	0.95\%	&	 bll 	&	 bll 	&	 bll 	&	 bll 	&	bll	&	$-$	&	bll	&	bll   	\\
3FGL J0043.5$-$0444	&	1.475	& 1.735	&	-14.170	&	0.023	&	-8.524	&	-9.726	&	-10.372	&	1.605	&	100.00\%	&	0.00\%	&	 bll 	&	 bll 	&	 bll 	&	 bll 	&	bll	&	bll	&	bll	&	bll   	\\
3FGL J0043.7$-$1117	&	1.397	& 1.594	&	-14.050	&	2.115	&	-9.092	&	-13.231	&	-10.386	&	1.442	&	100.00\%	&	0.00\%	&	 bll 	&	 bll 	&	 bll 	&	 bll 	&	bll	&	$-$	&	bll	&	bll   	\\
3FGL J0045.2$-$3704	&	2.518	& 2.543	&	-11.319	&	0.526	&	-7.845	&	-8.529	&	-11.132	&	2.240	&	3.19\%	&	96.81\%	&	 fsrq 	&	 fsrq 	&	 fsrq 	&	 fsrq 	&	fsrq	&	$-$	&	fsrq	&	fsrq     	\\
3FGL J0049.4$-$5401	&	2.292	& 2.143	&	-13.013	&	0.142	&	-8.368	&	-9.116	&	-10.561	&	1.653	&	99.40\%	&	0.60\%	&	 bll 	&	 bll 	&	 bll 	&	 bll 	&	fsrq	&	$-$	&	bll	&	bll   	\\
3FGL J0050.0$-$4458	&	2.526	& 2.528	&	-12.023	&	0.547	&	-8.269	&	-9.061	&	-14.818	&	1.836	&	0.61\%	&	99.39\%	&	 fsrq 	&	 fsrq 	&	 fsrq 	&	 fsrq 	&	fsrq	&	$-$	&	fsrq	&	unc	\\
3FGL J0051.2$-$6241	&	1.635	& 1.663	&	-13.074	&	1.834	&	-8.250	&	-8.943	&	-9.676	&	1.701	&	100.00\%	&	0.00\%	&	 bll 	&	 bll 	&	 bll 	&	 bll 	&	bll	&	$-$	&	bll	&	bll   	\\
3FGL J0055.2$-$1213	&	2.420	& 2.397	&	-12.466	&	0.604	&	-8.388	&	-8.907	&	-10.827	&	1.838	&	21.42\%	&	78.58\%	&	 fsrq 	&	 bll 	&	 bll 	&	 fsrq 	&	fsrq	&	$-$	&	fsrq	&	unc	\\
3FGL J0103.7$+$1323	&	1.716	& 1.984	&	-13.195	&	2.436	&	-8.719	&	-9.570	&	-10.742	&	1.722	&	99.21\%	&	0.79\%	&	 bll 	&	 bll 	&	 bll 	&	 bll 	&	bll	&	bll	&	bll	&	bll   	\\
3FGL J0107.0$-$1208	&	1.778	& 2.180	&	-12.943	&	0.964	&	-8.401	&	-9.149	&	-11.097	&	1.514	&	88.44\%	&	11.56\%	&	 bll 	&	 bll 	&	 bll 	&	 bll 	&	bll	&	$-$	&	bll	&	bll   	\\
3FGL J0116.2$-$2744	&	1.237	& 2.023	&	-13.369	&	1.034	&	-10.128	&	-9.161	&	-10.553	&	1.606	&	100.00\%	&	0.00\%	&	 bll 	&	 bll 	&	 bll 	&	 bll 	&	bll	&	$-$	&	bll	&	bll   	\\
3FGL J0121.7$+$5154	&	0.928	& 1.984	&	-13.406	&	0.437	&	-8.289	&	-9.269	&	-10.613	&	1.586	&	100.00\%	&	0.00\%	&	 bll 	&	 bll 	&	 bll 	&	 bll 	&	bll	&	$-$	&	bll	&	bll   	\\
3FGL J0127.2$+$0325	&	1.208	& 1.899	&	-12.793	&	1.603	&	-9.120	&	-8.783	&	-10.125	&	1.695	&	100.00\%	&	0.00\%	&	 bll 	&	 bll 	&	 bll 	&	 bll 	&	bll	&	$-$	&	bll	&	bll   	\\
3FGL J0132.5$-$0802	&	2.488	& 1.753	&	-13.863	&	1.681	&	-11.932	&	-12.006	&	-10.425	&	1.517	&	80.33\%	&	19.67\%	&	 bll 	&	 bll 	&	 bll 	&	 bll 	&	bll	&	$-$	&	bll	&	bll   	\\
3FGL J0133.2$-$5159	&	3.078	& 2.628	&	-12.079	&	0.722	&	-8.054	&	-9.077	&	-10.772	&	1.681	&	21.29\%	&	78.71\%	&	 fsrq 	&	 fsrq 	&	 fsrq 	&	 fsrq 	&	fsrq	&	$-$	&	fsrq	&	bll   	\\
3FGL J0133.3$+$4324	&	2.179	& 2.301	&	-12.602	&	1.617	&	-8.572	&	-8.777	&	-10.996	&	1.720	&	52.25\%	&	47.75\%	&	 bll 	&	 bll 	&	 bll 	&	 bll 	&	fsrq	&	$-$	&	unc	&	bll   	\\
3FGL J0134.5$+$2638	&	1.485	& 1.991	&	-12.750	&	3.036	&	-9.044	&	-8.804	&	-10.387	&	1.764	&	100.00\%	&	0.00\%	&	 bll 	&	 bll 	&	 bll 	&	 bll 	&	bll	&	$-$	&	bll	&	bll   	\\
3FGL J0139.9$+$8735	&	1.063	& 1.891	&	-13.833	&	1.268	&	-8.273	&	-9.975	&	-10.342	&	1.624	&	99.98\%	&	0.02\%	&	 bll 	&	 bll 	&	 bll 	&	 bll 	&	bll	&	$-$	&	bll	&	bll   	\\
\enddata
\tablecomments{Column 1 shows the 3FGL names; Column 2 shows the radio flux (log${\rm F_R}$); 
the $\gamma$-ray photon spectral index ($\Gamma_{\rm ph}$), flux density (log$F_D$), curve significance ($C_S$),
the integral  photon flux in 100 to 300 MeV (log$F_1$),  0.3 to 1 GeV (log$F_2$), 10 to 100 GeV (log$F_3$) and variability index (${\rm logVI}$) are listed in Columns 3-9 respectively.
Columns 10 and  11 show the probabilities $P_{Bi}$ and $P_{Fi}$ that a BCU $i$ belongs to the BL Lacs or FSRQs classes from $Mclust$ method, respectively.  
Columns  12 reports the optical classification ($\rm Class_{M}$) of BCUs using $Mclust$ method, where  ``{bll}" and ``{fsrq}"  indicate  BL Lac and  FSRQ, respectively. 
The classification of $Fermi$ BCUs using the Decision Trees (${\rm Class}_{DT}$), Random Forest (${\rm Class}_{RS}$) 
and Support Vector Machines (${\rm Class}_{SVM}$)  are reported in Columns 13, Columns 14 and Columns 15 respectively. 
Columns 16 ($Y$) and 17 ($M$) show the identified BL Lacs  and FSRQs reported in Yi et al. (2017, and references therein) by researching photon spectral index
and in Massaro et al. (2016, and references therein) by optical spectroscopic observations. 
Columns 18 ($LP$)  and Columns 19 ($Chi16$) 
list the classifications (``{unc}" for uncertain) in  Lefaucheur \& Pita (2017) using multivariate classifications,
and in Chiaro et al. (2016) using artificial neural networks (ANN) machine-learning techniques.}
Table \ref{DA_result} is published in its entirety in the machine-readable format. A portion is shown here for guidance regarding its form and content.
\end{deluxetable*}
\end{longrotatetable}

\end{document}